\documentclass[prb,preprint]{revtex4-1} 

\usepackage{amsmath}  
\usepackage{amsfonts} 
\usepackage{graphicx} 

\begin{document}

\title{Describing the hierarchical gravitational three-body problem with force and torque vectors}

\author{Barnab{\'a}s Deme}
\email{deme.barnabas@gmail.com}
\affiliation{Institute of Physics, E\"otv\"os University, P\'azm\'any P. s. 1/A, Budapest, 1117, Hungary}

\date{\today}

\begin{abstract}
We investigate the hierarchical gravitational three-body problem, in which a binary is perturbed by a distant object that orbits on a Keplerian ellipse around the binary itself. This phenomenon, known as Kozai-Lidov mechanism in the literature, results in large-amplitude oscillations of the orbital elements, like eccentricity, which may significantly influence the evolution of many astrophysical systems: from minor bodies and planets to even supermassive black holes. The standard approach found in the literature describes this phenomenon in the framework of Hamiltonian mechanics. Here we derive the long term evolution of the triple with elementary tools, treating the binary as a dipole interacting with the gravitational field of the tertiary and describe the dynamics using forces and torques instead of the usual Hamiltonian formalism. This highlights another aspect of the problem and is likely an easier way to introduce the problem at the undergraduate level. 
\end{abstract}

\maketitle

\section{Introduction} 

The gravitational three-body problem is one of the oldest open questions of physics.\cite{valtonen} In contrast to the two-body case, it is non-integrable since it lacks the sufficient number of first integrals. \cite{masoliver} As a consequence, instead of finding new first integrals, other attempts have been made to understand the long term evolution of the system. For example, recent efforts have focused on finding a statistical solution, i.e. to predict the distribution of the orbital elements after a strong three-body interaction.\cite{stone} 

Another approach, which has a longer history, is to make approximating assumptions about the three-body configuration, for which the long-term evolution may be explored analytically. One such assumption is to neglect the mass of one of the bodies, which results in the restricted three-body problem. The most famous discovery in this field is probably the prediction of the equilibrium Lagrange points.\cite{szebehely}

Besides the restricted configuration, three body systems in the hierarchical configuration also received attention in the past decades. These systems consist of a binary, which is perturbed by a \textit{distant} tertiary. The members of the binary constitute the inner binary, and their barycenter together with the tertiary forms the outer binary (see Fig.~\ref{configuration}). In this case, the ratio of the inner and outer orbital separations ($|\mathbf{r}|/|\mathbf{R}|=r/R$) is a small parameter.

The standard technique to describe the dynamics of such systems uses the Hamiltonian formalism.\cite{goldstein} 
One finds the perturbing Hamiltonian, expresses it with the canonical orbital elements (these are usually the so-called Delaunay elements) and expands it into a power series with respect to the small parameter. Depending on the desired accuracy, the Hamiltonian is truncated at a particular order: in a multipole expansion, when the terms smaller than $\mathcal{O}(r^2/R^2)$ are omitted, it is called the quadrupole approximation (we note, however, that the gravitational force is the minus gradient of the potential in the Hamiltonian, and hence is on the order of $\mathcal{O}(r/R)$ after differentiating with respect to $\mathbf{r}$). The long-term evolution of the system may be derived by time-averaging the perturbation over the inner and outer orbital periods. The equations of motion are then obtained from this double-averaged Hamiltonian. It turns out that sufficiently inclined triples (for which the $i$ angle between the inner and outer orbital planes is sufficiently large) are subject to large-amplitude eccentricity and inclination oscillations. This phenomenon is called the Kozai-Lidov (KL) mechanism,\cite{zeipel,lidov,kozai,ito_ohtsuka} which plays an important role in the dynamics of several astrophysical phenomena: from the formation of hot Jupiters to the merger of black holes.\cite{naoz}

Although the Hamiltonian formalism is elegant and easy-to-use, it is abstract in the sense that it does not highlight the role of force and torque vectors. Even if we express the Hamiltonian with the angular momentum and eccentricity vectors,\cite{katz,rungelenzlaplace} it is non-trivial how the KL mechanism follows from the basic equations of translational ($m\ddot{\mathbf{r}}=\mathbf{F}$) and rotational ($m(\mathbf{r}\times\ddot{\mathbf{r}})=\mathbf{r}\times \mathbf{F}$) motions. The aim of this paper is to derive the KL mechanism from these expressions. Up to the knowledge of the author, this approach is new in the literature.

The structure of the paper is as follows. In Sec. II we derive the force and torque vectors of the KL mechanism. In Sec. III we discuss the most important physical implications and give a qualitative picture of the dynamics.

Throughout the paper, the gravitational constant is set to 1. The scalar and vectorial products of $\mathbf{a}$ and $\mathbf{b}$ are denoted by $\mathbf{a}\mathbf{b}$ and $\mathbf{a}\times\mathbf{b}$, respectively.

\section{The KL force and torque}

\begin{figure}
\centering
\includegraphics[scale=0.6]{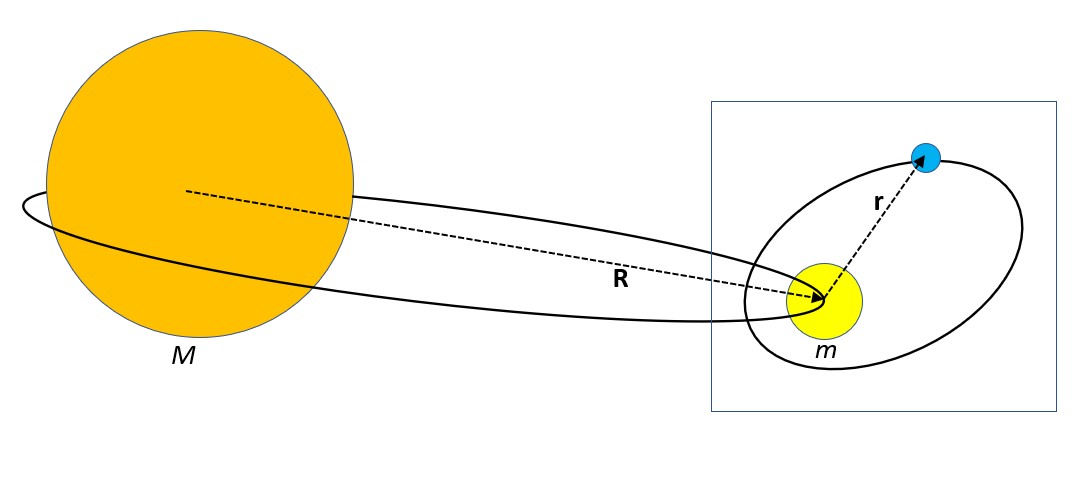}\\
\includegraphics[scale=0.6]{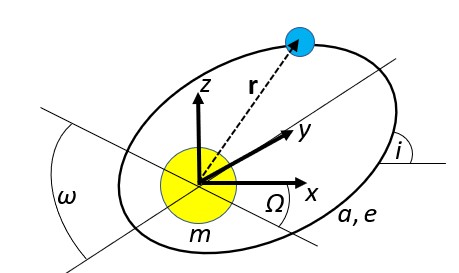}
\caption{\textit{Top:} The hierarchical three-body problem, in which $|\mathbf{r}| \ll |\mathbf{R}|$ (the figure is not to scale). \textit{Bottom:} The orbital elements of the inner binary from the blue box in the upper panel. The reference plane to which they are defined is the orbital plane of the outer binary.}
\label{configuration}
\end{figure}
In what follows we consider a three-body configuration, in which the ratio of the separations is small, hence we refer to it as hierarchical (see Fig.~\ref{configuration}). Two of the bodies, which we call primary and secondary, comprise the inner binary. Let us measure mass in units of the primary mass. The secondary (with mass $1 \ll m$) is assumed to orbit on a circular orbit around the tertiary ($M$), which is also much more massive ($m \ll M$). For instance, this may represent a star-planet-moon system so we refer to the objects as star, planet and moon hereafter. The goal of this work is to describe the long-term, i.e. \textit{secular} evolution of the inner binary (the moon around the planet), which is perturbed by the star. We use the quadrupole approximation, i.e., we keep only those terms in the expression of the force which are $\mathcal{O}(r/R)$ (or $\mathcal{O}(r^2/R^2)$ in the expression of the potential/Hamiltonian).

The validity of these assumptions can be confirmed by the example of the Sun-Earth-Moon system.\cite{beletsky} The ratio of the inner and outer separations is $r/R\approx a_\mathrm{Moon}/a_\mathrm{Earth}\approx 3\times 10^{-3}$, so the system is hierarchical enough to omit the next term in the Hamiltonian after the quadrupole (as that is $\sim10^{-9}$). If the mass of the Moon is set to 1, then $1\ll m_\mathrm{Earth}\approx 80$ and $m_\mathrm{Earth} \ll M_\mathrm{Sun}\approx 2.7\times 10^7$, so the orders of magnitudes assumed for the masses are also correct. Moreover, as $e_\mathrm{Earth}\approx 0.017$, the assumption of a circular outer orbit is valid, too. We note that the assumptions above (except hierarchicity) may be left and the results below can be generalized in a straightforward way.

Let us examine the behaviour of the orbital angular momentum vector, which is
\begin{equation}
    \mathbf{J}=J\left(
    \begin{array}{c}
            \sin i \sin \Omega\\
            -\sin i \cos \Omega\\
            \cos i\\
    \end{array}\right),
\end{equation}
where $J=\sqrt{ma(1-e^2)}$ (for the notations see Fig.~\ref{configuration}). Orbital elements hereafter refer to the inner (planet-moon) binary. We note here that the spins of the bodies are not taken into account, because their relative contributions to the angular momenta are negligible (the spin to orbital angular momentum ratios are $2\times10^{-5}$ and $7\times10^{-7}$ for the Moon and Earth, respectively). The evolution of the angular momentum is driven by a torque, $\dot{\mathbf{J}}=\mathbf{T}=\mathbf{r}\times \mathbf{F}$, where the $\mathbf{r}$ vector from the planet to the moon acts as the lever arm vector and 
\begin{equation}
    \mathbf{F}=-\frac{m}{r^3}\mathbf{r}+\mathbf{F}_\mathrm{out}.
\end{equation}
The first term on the right-hand side is the gravitational attraction on the moon exerted by the planet, while the second is that on the moon by the star: 
\begin{equation}
    \mathbf{F}_\mathrm{out}= M \frac{3\mathbf{R}(\mathbf{R}\mathbf{r})-R^2\mathbf{r}}{R^5}.
\end{equation}
This is completely analagous to the Coulomb force exerted on an electromagnetic dipole, where $\mathbf{r}$ plays the role of the dipole vector.\cite{jackson} Its physical motivation is that the inner (planet--moon) binary cannot be treated to be point-like, because it has a finite, although small extent ($r \sim a \ll R$) which can be modelled as a dipole in the first approximation in a multipole expansion.

Now let us calculate the work this force does on the moon's orbit. We are not interested in those effects which happen within one orbital period, so we sum up the $\mathbf{F}\mbox{ d}\mathbf{r}$ work elements along one orbit:
\begin{equation}\label{workofforce}
    W=\oint \mathbf{F}\mbox{ d}\mathbf{r}=\oint \mathbf{F} \dot{\mathbf{r}}\mbox{ d}t = \oint \left( -\frac{m}{r^3}\mathbf{r}+\mathbf{F}_\mathrm{out} \right)\dot{\mathbf{r}} \frac{1-e\cos E}{m^{1/2}a^{-3/2}}\mbox{ d}E,
\end{equation}
where we used $\mathrm{d}t=\mathrm{d}l/(m^{1/2}a^{-3/2})$ and  $\mathrm{d}l=1-e\cos E\mbox{ d}E$ from the Kepler equation ($l$ and $E$ are the so-called mean and eccentric anomalies, respectively; see Appendix \ref{work} for the technical details). In order to evaluate the integral we may use that
\begin{equation}
    \mathbf{r}=\left(
    \begin{array}{c}
            x\\
            y\\
            z\\
    \end{array}\right)=\mathbf{R}_z(\Omega)\mathbf{R}_x(i)\mathbf{R}_z(\omega)\left(
    \begin{array}{c}
            a(\cos E - e)\\
            a\sqrt{1-e^2} \sin E\\
            0\\
    \end{array}\right),
\end{equation}
and
\begin{equation}
    \dot{\mathbf{r}}=\left(
    \begin{array}{c}
            \dot{x}\\
            \dot{y}\\
            \dot{z}\\
    \end{array}\right)=\mathbf{R}_z(\Omega)\mathbf{R}_x(i)\mathbf{R}_z(\omega)\frac{m^{1/2}a^{-3/2}}{1-e\cos E}\left(
    \begin{array}{c}
            -a\sin E\\
            a\sqrt{1-e^2} \cos E\\
            0\\
    \end{array}\right),
\end{equation}
where $\mathbf{R}_j(\alpha)$ is the rotation matrix around axis $j$ by an angle $\alpha$.

Calculating the integral in Eq.~\eqref{workofforce} is not complicated and eventually gives zero: $W=0$. It is not surprising, because we know from first principles that the loop integral of any force that is the gradient of a potential is zero. In other words, apart from some small periodic oscillations on the orbital timescale, the energy of the moon remains secularly constant. We note that the integration above can be considered as an averaging with respect to the mean anomaly $l$, hence we can also write
\begin{equation}\label{averagework}
    \langle W \rangle_l =0.
\end{equation}

The torque is 
\begin{equation}
    \mathbf{T}=\mathbf{r}\times\mathbf{F} = M\mathbf{r}\times \frac{3\mathbf{R}(\mathbf{R}\mathbf{r})-R^2\mathbf{r}}{R^5}.
\end{equation}

In order to understand how this torque drives the long-term evolution of the planet-moon binary, we calculate its averaged value over one orbital period. In doing so, similarly to what we did in Eq.~\eqref{workofforce}, we sum up the small $\mathbf{T}\mbox{ d}l$ elements to get the net torque that affects the dynamics of the moon over several orbital times. However, unlike the case of the mechanical work, the integration over the inner (planet--moon) orbit now yields a non-vanishing torque, so we extend it to the outer orbit, too:
\begin{equation}\label{doubleaveragedtorque}
    \langle \langle \mathbf{T} \rangle_{\phi} \rangle_{l} = \Tilde{\mathbf{T}}=\frac{3M}{8R^3}a^2 \sin i \left(
    \begin{array}{c}
            \cos i \left( -2-3e^2+5e^2\cos(2\Omega) \right)\cos \Omega -5e^2\sin(2\omega)\sin\Omega\\
            \cos i \left( -2-3e^2+5e^2\cos(2\Omega) \right)\sin \Omega +5e^2\sin(2\omega)\cos\Omega\\
            0\\
    \end{array}\right),
\end{equation}
where $\phi$ is the mean anomaly of the outer orbit. Just like previously, $\langle...\rangle_{l}=\int_0^{2\pi} ... (1-e \cos E)\mbox{ d}E/(2\pi)$ is the averaging over the moon's orbit around the planet, while $\langle...\rangle_{\phi}=\int_0^{2\pi} ... \mbox{ d}\phi/(2\pi)$ is the one over the orbit of the planet-moon binary itself around the star. The reason for why the latter integral does not have a Jacobian determinant similar to $1-e\cos E$ is that the outer orbit (planet around the star) is circular.

In order to grasp the geometrical meaning of the vector in Eq.~\eqref{doubleaveragedtorque}, we decompose it into two components. We first examine the parallel projection of $\Tilde{\mathbf{T}}$ onto $\hat{\mathbf{e}}_{\mathbf{J}}$, the unit vector in the direction of the inner (planet--moon) binary's angular momentum. The parallel component, which changes only the magnitude of the angular momentum but not its direction, is 
\begin{equation}\label{paralleltorquevector}
    \mathbf{T}_{\parallel}=\hat{\mathbf{e}}_{\mathbf{J}} \left(\hat{\mathbf{e}}_{\mathbf{J}} \Tilde{\mathbf{T}}\right) = -\frac{15M}{8R^3}a^2e^2\sin^2 i \sin(2\omega) \hat{\mathbf{e}}_{\mathbf{J}}.
\end{equation}
The direction of $\mathbf{J}$ is driven by the perpendicular component:
\begin{equation}\label{perpendicular1}
    \mathbf{T}_{\bot}= \Tilde{\mathbf{T}}- \hat{\mathbf{e}}_{\mathbf{J}}  \left(\hat{\mathbf{e}}_{\mathbf{J}}\Tilde{\mathbf{T}}\right),
\end{equation}
which leaves the magnitude of $\mathbf{J}$ unchanged. The components of this vector are 
\begin{equation}\label{torquex}
    (\mathbf{T}_{\bot})_x=-\frac{3a^2M\sin(2i)}{16R^3}\left((2+3e^2-5e^2\cos (2\omega))\cos \Omega+5e^2\cos i \sin (2\omega)\sin \Omega\right),
\end{equation}
\begin{multline}\label{torquey}
    (\mathbf{T}_{\bot})_y=\frac{3a^2M\sin(2i)}{16R^3}\left((-2-3e^2-5e^2\cos (2\omega))\sin \Omega+5e^2\cos i \sin (2\omega)\cos \Omega\right),
\end{multline}
\begin{equation}\label{torquez}
    (\mathbf{T}_{\bot})_z=\frac{15a^2e^2M\sin(2i)\sin i\sin(2\omega)}{16R^3}.
\end{equation}

\section{Discussion}

The dynamics governed by the forces and torques above has some remarkable features. First, the inner semi-major axis is constant. It follows from the fact that the work of the tertiary on the inner binary is zero (see Eq.~\eqref{averagework}), consequently the total orbital energy ($h$) is secularly conserved. As $a=-m/(2h)$, the semi-major axis remains constant, too.  

Secondly, $\langle\langle T_z \rangle\rangle$ turns out to be zero (see Eq.~\eqref{doubleaveragedtorque}), which means that
\begin{equation}\label{happycoincidence}
    \dot{J}_{1,z}=0 \to \sqrt{m a (1-e^2)}\cos i = \mathrm{const.} \to \sqrt{1-e^2}\cos i = \mathrm{const.},
\end{equation}
where we took into account that $a=\mathrm{const.}$ The last expression is known as the Kozai constant, while the vanishing of the vertical torque is coined the 'happy coincidence' in the Hamiltonian framework (because it makes the system integrable).\cite{lidov76}

\begin{figure}
\centering
\includegraphics[scale=0.4]{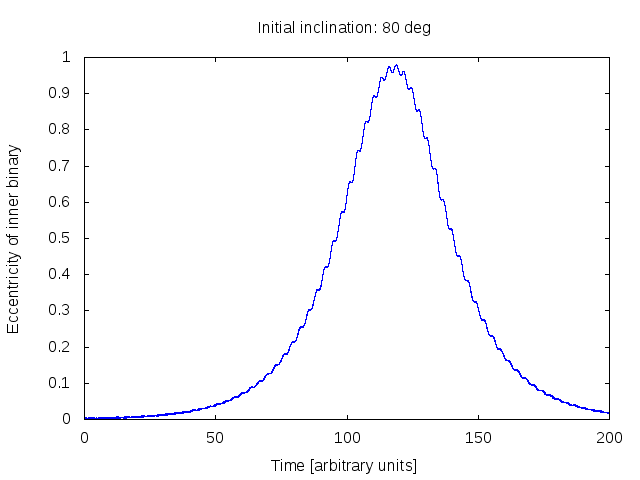}
\includegraphics[scale=0.4]{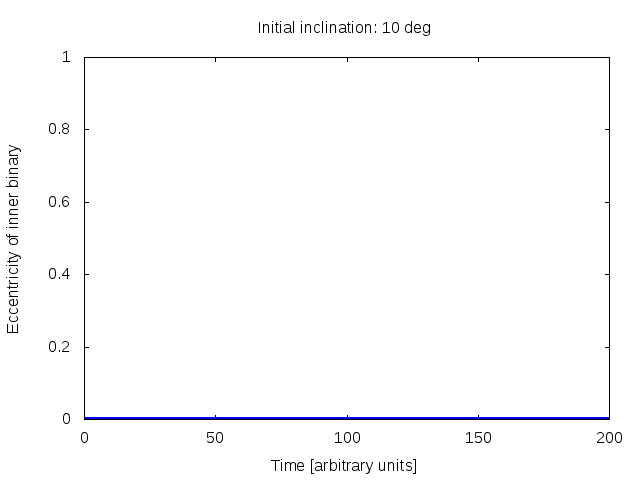}\\
\includegraphics[scale=0.4]{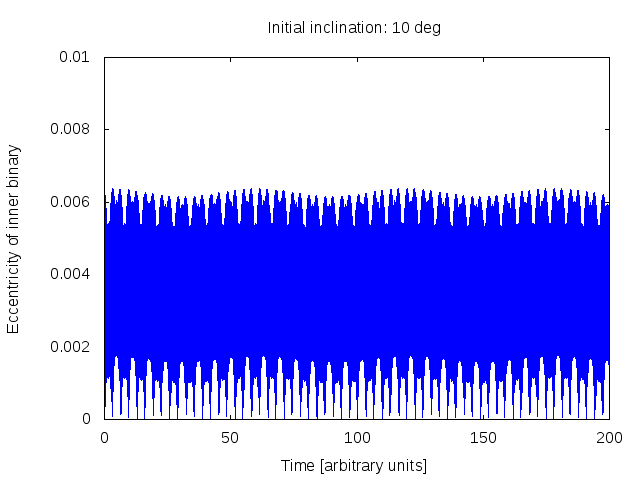}
\caption{Numerical integration of a hierarchical triple system with $M=10^6$ and $m=10^3$ and with initial parameters $a_1=1$, $R=100$ and $e=0.0$ (arbitrary units with the gravitational constant $G=1$). \textit{Top left:} The initial inclination is $i=80^\circ$. The initially zero inner eccentricity is pumped up close to unity, i.e., to a very elongated orbit. \textit{Top right:} The same set-up but with $i=10^\circ$ initially. \textit{Bottom:} The same as in the top right panel, but with a 100× zoom-in on the $y$-axis.}
\label{oscillations}
\end{figure}

Thirdly, unlike the orbital energy, the inner angular momentum does not remain constant but varies with time, because $\tilde{\mathbf{T}}\neq \mathbf{0}$. As $a=\mathrm{const.}$, the evolution of its magnitude ($J=\sqrt{ma(1-e^2)}$) is identical to that of the inner eccentricity. Through $\sin(2\omega)$, it is subject to a periodic driver,\cite{kozairesonance} which becomes more efficient with increasing inclination (see Eq.~\eqref{paralleltorquevector}). In other words, the eccentricity of highly inclined orbits oscillate. This is demonstrated in Fig.~\ref{oscillations} by numerical simulations. The top left panel shows the case initialized with $i=80^\circ$: the KL mechanism is very effective, it enhances the eccentricity significantly. The top right panel exhibits the opposite case, where KL is very weak due to the low initial inclination ($i=10^\circ$). The bottom panel is a zoom-in version of the top right. The small-amplitude oscillations superposed on the long-term curves are the results of the inner and outer orbital motions that we ignore by the orbit-averaging in Eq.~\eqref{doubleaveragedtorque}. Note that the inclination itself also changes in accord with the eccentricity, because the two are related by the Kozai constant (see Eq.~\eqref{happycoincidence}). 

The timescale of the oscillation of the orbital elements can be easily estimated. The subject of the KL effect is the angular momentum: the rate of its change is given by the double-averaged torque, see Eq.~\eqref{doubleaveragedtorque}. The characteristic timescale is given by $t_\mathrm{KL}\approx J/\dot{J}\approx J/\tilde{T}$. Since the orbital elements (except $a$) vary with time due to the torque, they affect the exact value of the timescale. However, $e\in [0,1]$ and the angles are in the arguments of sine and cosine functions, so they induce variations only of order unity and can be omitted in an order-of-magnitude formula. Hence
\begin{equation}
    t_\mathrm{KL}\approx\frac{J}{\tilde{T}}\approx\frac{\sqrt{ma}}{Ma^2/R^3}=\frac{m^{1/2}R^3}{Ma^{3/2}}.
\end{equation}
It is much longer than the outer orbital time, $t_\mathrm{out}\approx M^{-1/2}R^{3/2} $ (i.e. it is secular), because
\begin{equation}
    \frac{t_\mathrm{KL}}{t_\mathrm{outer}}=\left(\frac{m}{M}\right)^{1/2}\left(\frac{R}{a}\right)^{3/2}
\end{equation}
and $1\ll R/a$.

In conclusion, the KL mechanism is a secular phenomenon in the hierarchical gravitational three-body problem that keeps the semi-major axis constant but makes the eccentricity and the inclination oscillate. The amplitude of eccentricity oscillation is most prominent if the perturbing outer object orbits in a perpendicular plane with respect to the binary.

We note here that the torque in Eq.~\eqref{doubleaveragedtorque} is not responsible for the evolution of $\omega$ (the periapsis may precess even in the absence of torques). The quantity that triggers the variation of $\omega$ could be the time-derivative of the Runge-Lenz vector, just like the torque is the time-derivative of the angular momentum vector.

\appendix*

\section{The Kepler equation}\label{work}

The Kepler equation expresses the connection between the eccentric ($E$) and mean ($l$) anomalies:
\begin{equation}\label{keplereq}
    E-e\sin E=l,
\end{equation}
where $l$ is proportional to the time:
\begin{equation}
    l=l_0+m^{1/2}a^{-3/2}t.
\end{equation}
Differentiating Eq.~\eqref{keplereq} with respect to time yields
\begin{equation}
    (1-e\cos E)\dot{E}=m^{1/2}a^{-3/2},
\end{equation}
which we can rearrange as
\begin{equation}
    \mathrm{d}E=\frac{m^{1/2}a^{-3/2}}{1-e\cos E}\mathrm{ d}t.
\end{equation}

\begin{acknowledgments}

I thank Bence Kocsis for stimulating discussions. This project has received funding from the European Research Council (ERC) under the European Union's Horizon 2020 research and innovation programme under grant agreement No 638435 (GalNUC) and by the Hungarian National Research, Development, and Innovation Office grant NKFIH KH-125675.
\end{acknowledgments}


\begin{thebibliography}{99}
\bibitem{kozairesonance} We note that although $\omega$ generally rotates across the $[0,2\pi]$ interval, the more detailed calculation reveals that it may be tuned to librate around $\pm \pi/2$. This is called the Kozai resonance. Nevertheless, this fact does not alter our qualitative picture about the periodic nature of $\mathbf{T}_{\parallel}$.
\end{thebibliography}
\end{document}